\documentclass[10pt,journal,compsoc]{IEEEtran}

\usepackage[T1]{fontenc}
\usepackage{graphicx}
\usepackage{amsmath,amssymb,amsfonts}
\usepackage{algorithmic}
\usepackage{textcomp}
\usepackage{xcolor}
\usepackage{autobreak}
\usepackage{multirow}
\usepackage{array}
\usepackage{lscape}
\usepackage{rotating}
\usepackage{booktabs}

\usepackage{tikz}

\usepackage{longtable}
\newcommand{\tabincell}[2]{\begin{tabular}{@{}#1@{}}#2\end{tabular}}

\newcommand\Future{%
        \mathrel{\text{\raisebox{.7ex}{\tikz[baseline] \draw (0ex,1ex) -- (-1ex,0ex) (-1ex,0ex) -- (0ex,-1ex) (0ex,-1ex) -- (1ex,0ex) (1ex,0ex) -- (0ex,1ex);}%
}}}
\newcommand\Global{%
        \mathrel{\text{\raisebox{.7ex}{\tikz[baseline] \draw (0.707ex,0.707ex) -- (-0.707ex,0.707ex) (-0.707ex,0.707ex) -- (-0.707ex,-0.707ex) (-0.707ex,-0.707ex) -- (0.707ex,-0.707ex) (0.707ex,-0.707ex) -- (0.707ex,0.707ex);}%
}}}
\newcommand\Next{%
        \mathrel{\text{\raisebox{.7ex}{\tikz[baseline] \draw (0,0) circle (1ex);}%
}}}

\ifCLASSINFOpdf
\else
\fi
\begin{document}

%
\title{Vulnerabilities Analysis and Secure Controlling for Unmanned Aerial System Based on Reactive Synthesis}
%
%
%
%

\author{Dong~Yang,
        Wei~Dong,
        Wei~Lu,
        Yanqi~Dong,
        and~Sirui~Liu 
\IEEEcompsocitemizethanks{\IEEEcompsocthanksitem D. Yang, W. Dong, W. Lu, S. Liu and Y. Dong was with the 
with School of Computer, National University of Defense Technology, Hunan, China, 410074.\protect\\  
E-mail: \{yangdong15, wdong, luwei09, d16945, liusirui\}@nudt.edu.cn}
\thanks{Manuscript received April 19, 2005; revised August 26, 2015.}}

\markboth{Journal of \LaTeX\ Class Files,~Vol.~14, No.~8, August~2015}%
{Shell \MakeLowercase{\textit{et al.}}: Bare Advanced Demo of IEEEtran.cls for IEEE Computer Society Journals}
%




\IEEEtitleabstractindextext{%
\begin{abstract}
  Complex Cyber-Physical System (CPS) such as Unmanned Aerial System (UAS) got rapid development these years, but also became vulnerable to GPS spoofing, packets injection, 
  buffer-overflow and other malicious attacks. Ensuring the behaviors of UAS always keeping secure no matter how the environment changes, would be a prospective direction for
  UAS security. This paper aims at introducing a patterns-based framework to describe the security properties of UAS, and presenting a reactive synthesis-based approach to implement
  the automatic generation of secure UAS controller. First, we study the operating mechanism of UAS and construct a high-level model consisting of actuator and monitor. Besides, we 
  analyze the security threats of UAS from the perspective of hardware, software and cyber physics, and then summarize the corresponding specification patterns of security properties 
  with LTL formulas. With the UAS model and security specification patterns, automatons for controller can be constructed by General Reactivity of Rank 1 (GR(1)) synthesis algorithm, 
  which is a two-player game process between Unmanned Aerial Vehicle (UAV) and its environment.
  Finally, we experimented under the Ardupilot simulation platform to test the effectiveness of our method.
  
  \end{abstract}

  }
    
    \maketitle

  \section{Introduction}
  With the rapid development of technologies, people will focus more on the CPS area in the future. Unmanned Aerial System, an Artificial Intelligence based complex CPS, which deeply integrates the technologies 
  of environmental awareness, data analysis, authentication and heterogeneous networks, has already been widely used in different fields. According to the application scenario, UAS can usually be divided into
   three main categories: military drones, industrial drones, and commercial drones. Today, UAS plays an important role in the deployment of cooperative engagement, reconnaissance, remote sensing, aerial photography, 
   agroforestry and Smart City service, because of the advantages of small-size, low-cost, high-speed, convenience and good flexibility, etc.
  
  Along with the increasing development, the security threats of UAS are also increasing. Even with the integration of advanced technologies, these CPSs are still prone to faults due to unpredicted state transitions 
  and external interference. Attackers can implement the attacks through the hardware, software, or network of system, to compromise the confidentiality, integrity and availability, such as malicious 
  injection, authentication bypass, GPS spoofing and DDoS attack, etc.
  For example, the unencrypted real-time video signal transmission of MQ-1 Predator resulted in a video feed interception by Iranian militants in 2009. A GPS spoofing attack was performed by Iranian forces on RQ-170,
   resulting in a successful capture on American UAV in 2011. The Ground Control Station (GCS) of Creech Air Force Base in Nevada has been infected by a computer virus named ``keylogger'' 
   in 2011\cite{8088163}. In recent years, major security conferences and competitions are devoted to the researches regarding attack-defense on products from commercial drone manufacturers such as Parrot and 
   DJI, and various vulnerabilities are exposed.
  
  Yet, CPS security, especially UAS security is more severe than conventional security. Complex network architectures, flexible physical environment, and excessive access interfaces, may lead to more vulnerabilities 
  and attack interfaces when manufacturers try to improve the quality of drones. Attackers can exploit the vulnerabilities to cause the sensitive data leaked, the system hijacked and even the drones crashed instantly.
  
  Reactive synthesis is a methodology about synthesizing a correct-by-construc-tion reactive system automatically, from given formal specifications. The obtained system is usually 
  represented in the form of automaton satisfying the specifications. The input of the automaton can be viewed as the environmental virables from sensors of robots, and the output is the actions performed by the
   actuators. The synthesis algorithm introduced in this paper is based on the GR(1) game\cite{DBLP:journals/jcss/BloemJPPS12}.

   In this paper, we design a patterns-based framework to describe the security properties of UAS, expand the work in\cite{5238617} and introduce a novel approach implementing the automatic generation 
   of the secure controller for UAS with GR(1) reactive synthesis. The contributions can be summarized as follows:
   1) we study the operating mechanism of UAS and abstract a high-level model of UAS with two components \textit{actuator} and \textit{monitor};
   2) we study the vulnerabilities and security properties of UAS, so as to extract the security specification patterns according to our requirements. The model and specifications are described in Linear Temporal
    Logic (LTL)\cite{Emerson90Temporal};
   3) we implement the automatic generation of the secure controller for UAS with the GR(1) game-based reactive synthesis algorithm applied in multi-UAV systems;

  \section{Preliminaries}
  \label{sec:Pre}
  LTL \cite{Emerson90Temporal} is widely used to describe specifications in formal method field. Syntactically, LTL extend propositional logic with the following temporal operators: $\Next$ (next), $\mathbf{U}$(until),
   $\Future$ (eventually), $\Global$ (globally). Semantically, given an infinite sequence $\pi$ composed by subsets of atom positions $AP$ and a position $i\in\mathbb{N}$, $\pi,i\models \varphi$ denotes that the 
   LTL formula $\varphi$ holds at the $i$-th position of $\pi$.
  
  To solve general LTL synthesis problems, a necessary two-phase translation will result in a double-exponential state blow-up. To alleviate computational complexity into an acceptable range, 
  a special restriction of LTL called GR(1) specification is taken into consideration \cite{Piterman06Synthesis}.
  
  For a system, its GR(1) specification consists of the following elements:
  \begin{itemize}
  \item $\varphi^e_i$ and $\varphi^s_i$ are propositional logic formulas without temporal operators, which are defined on $X$ and $Y$ respectively. They describe initial conditions of the behaviors of environment 
  and system respectively.
  \item $\varphi^e_t$ is a conjunction of several subformulas in the form of $\Global A_i$, where $A_i$ is a boolean formula defined on $X\cup Y\cup\Next X$, where $\Next X=\{\Next x\ |\ x\in X\}$. $\varphi^e_t$ 
  limits the relation between the next behaviors of environment and current state.
  \item $\varphi^s_t$ is a conjunction of several subformulas in the form of $\Global A_j$, where $A_j$ is a boolean formula defined on $X\cup Y\cup\Next X\cup\Next Y$, and $\Next Y=\{\Next y\ |\ y\in Y\}$. 
  $\varphi^s_t$ limits the relation between the next behaviors of the system and current state, as well as the next behaviors of environment.
  \item $\varphi^e_g$ and $\varphi^s_g$ are the conjunctions of several subformulas in the form of $\Global\,\Future B_i$, where $B_i$ is boolean formula defined on $X\cup Y$. They describe final goals of environment 
  and system respectively.
  \end{itemize}
  The formula $\phi=\varphi^e_g\Rightarrow\varphi^s_g$ is called a GR(1) formula \cite{Piterman06Synthesis}. Intuitively speaking, assumptions including $\varphi^e_i$, $\varphi^e_t$ and $\varphi^e_g$ constrain the 
  possible environments and guarantees including $\varphi^s_i$, $\varphi^s_t$ and $\varphi^s_g$ limit the system's behaviors, so the specification specifies the rules of $r_i$'s behaviors under given environments.
  
  The expression ability of GR(1) specification is strictly weaker than LTL (for example, $p\,\mathbf{U} \,q$ cannot be expressed by the fragment). However, most properties in practical, especially in reactive 
  systems, can be expressed by the restriction~\cite{Piterman06Synthesis}.
  
  Given a GR(1) specification, the synthesis algorithm to construct an automaton to satisfy the specification can be done in $O(n^3)$ time, where $n$ is the scale of the state 
  space \cite{Piterman06Synthesis}. The synthesis algorithm of GR(1) specifications is mainly a process of solving the game between the environment and system\cite{Piterman06Synthesis}. The controller
   synthesized by the algorithm is represented in the form of an automaton. According to the synthesized controller, when an admissible input sequence that satisfies environment 
   assumptions is given, the discrete path of the system can be acquired, which can guide the system to choose a position to go and activate/deactivate the corresponding actions.

  \section{Threat Analysis of Unmanned Aerial System}
  \subsection{System Structure of UAS}
  UAS can be generally divided into two parts of UAV and GCS. GCS can ba a remote controller(RC), smart mobile device, computer or even a military base, which establishes a bidirectional communication with UAV to 
  implement the remote state monitoring and motion controlling. UAV has various types such as fixed wing craft, rotor craft and so on. In addition to being operated remotely by humans, UAV can be autonomous or 
  semi-autonomous too. Autonomous UAV can sence from the environment with active or passive perception system, make a decision based on the real-time mission planning algorithm and command the actuators to execute 
  some specific behaviors to achieve disired goals.
  According to the function, most UAS consists of 4 layers shown in \figurename\ {\ref{System Structure of UAS}}:
  
  \begin{figure*}[htbp]
    \centering
    \includegraphics[width= 14 cm]{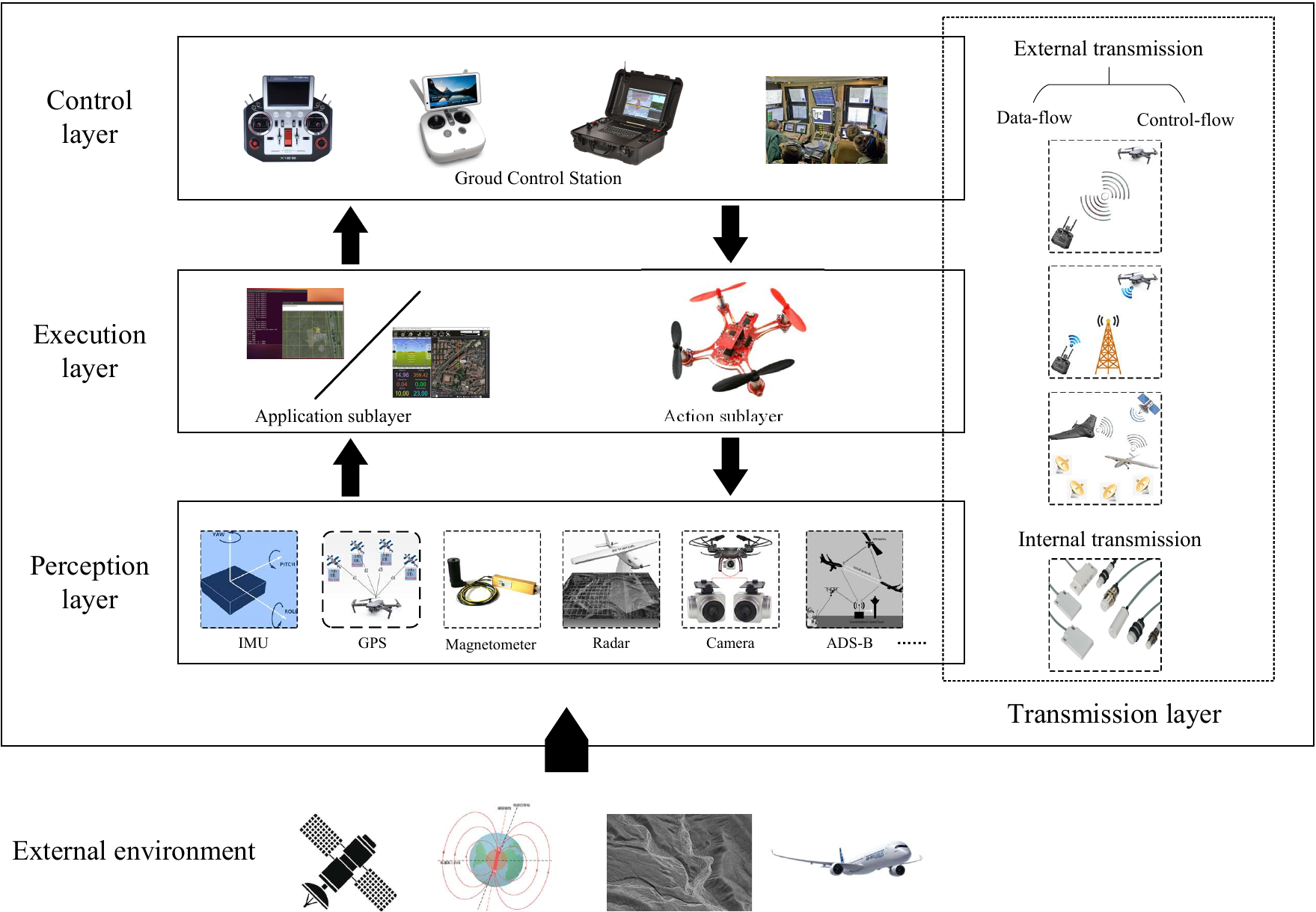}
    \caption{System Structure of UAS}\label{System Structure of UAS}
  \end{figure*}
  
  \begin{itemize}
  \item Perception layer. UAV is equipped with a variety of sensors: positioning and navigation is usually implemented by Global Navigation Satellite System(GNSS) and Inertial Measurement Unit(IMU), the former for 
  precise positioning and the latter for relative positioning and self-attitude perception; radar and camera are responsible for perceiving the surrounding environment and objects; Automatic Dependent 
  Surveillance-Broadcast(ADS-B) system can broadcast UAV's flight state and receive from other crafts. Multi-sensor fusion can effectively model its environment and detect sensor faults by checking the 
  characteristics of the state estimates’ statistical properties\cite{liu2013multi}.
  \item Execution layer. It consists of application sublayer and action sublayer. The onboard control system receives incoming data from perception layer and control layer, and generates appropriate commands to the 
  action sublayer. Action sublayer is actually a low-level and continuous controller for actuators of UAV. It gets the commands and drives the physical actuators to execute the dynamic motions. The current main 
  control systems are APM and PIXHawk, and the current main control algorithms are Strapdown Inertial Navigation, Kalman filter algorithm and PID algorithm.
  \item Control layer. GCS is the control center of UAS. It focuses on monitoring the flight state, load status and video data of the UAV and displays to the operators. Further more, it can make short term decisions 
  and long term planning based on the control algorithms and send the commands to the execution layer to achieve the flight planning, task scheduling and data management. GCS consists usually of 4 modules of 
  communication, information display, data storage and video playback.
  \item Transmission layer. UAV can construct various networks with different devices and nodes to accomplish the function of remote controlling, data transmission and real-time communication such as 
  air-to-ground(ATG), air-to-air(ATA) and satellite-data links, etc.
  \end{itemize}

  

  \subsection{Specification Patterns for UAS}\label{threats}
  UAS vulnerabilities involve extensive fields, such as sense, execution, control and data transmission. We design a pattern-based framework to describe their safety/security properties. 
  With the framework we can synthesize the corresponding secure controller against certain threats. A specification 
  pattern consists of 5 sections \cite{2020Synthesizing}:

  \begin{itemize}
    \item \textbf{Input} is the userdefined proposition that provide the information of the environment and the UAS, namely, the interface between real world and the abstract model of UAS.
     In a real control software of UAS (e.g., Ardupilot), the input can be the events from the flight logfiles.
   \item \textbf{Property} is the requirements of UAS that restrain the behaviors of system, usually expressed in \textbf{shall}/\textbf{shall not} statements.
   \item \textbf{Specification} is the formal description of the properties in LTL/MTL formulas.
   \item \textbf{Output} is the output of monitor generated by the property specification, can be used as the input of environment constraint ($\varphi^{e}$) of the GR(1) systhesizer.
   \item \textbf{Countermeasure} is the set of \textit{actions} UAS should take when the properties are violated, can be used as the output of system constraint ($\varphi^{s}$) of the GR(1) systhesizer.
  \end{itemize}
  
  Therefore, we analyze the threats and summarize the specification patterns for their corresponding properties: we introduce one possible  countermeasure and introduce its property specification based on the
   protective mechanism. The definitions and descriptions of input and output interfaces of UAS are shown in Tab. \ref{interface}.

  \begin{table*}[!htbp]
    \begin{center}
    \caption{Definition and description of input , output and countermeasure interface of UAS}
      \label{interface}
    \begin{tabular}{lll}
      \toprule
      \multicolumn{1}{c}{\textbf{Interface Type}} & \multicolumn{1}{c}{\textbf{Name}} & \multicolumn{1}{c}{\textbf{Description}}\\
      \toprule
      \multirow{20}*{Input} & {$ComponentFingerprint$} &  {Device fingerprint identification of internal physical components in the system}\\
      & {$UserID$} &  {The state of a user's identity to access the system, e.g.,account password, dongle, etc.}\\
      & {$FusionEstimation$} &  {System state estimation based on data fusion mechanism}\\
      & {$CmdGet$} &  {The system receives a control instruction}\\
      & {$ParmLength$} &  {The state of the length of each field in the instruction}\\
      & {$CPUuse$} &  {The state of CPU usage in the system}\\
      & {$MEMuse$} &  {The state of memory usage in the system}\\
      & {$BatLevel$} &  {The state of battery consumption in the system}\\
      & {$Mode$} &  {The system is in a specific mode state}\\
      & {$AbnormalCmd$} &  {The system has received an abnormal control instruction}\\
      & {$TakeOff$} &  {UAS receives the take-off command}\\
      & {$TAKEOFF$} &  {UAS is in TAKEOFF MODE}\\
      & {$landon$} &  {UAS receives the land-on command}\\
      & {$LandOn$} &  {UAV has completed landing}\\
      & {$LANDON$} &  {UAS is in LANDON MODE}\\
      & {$DeauthCmd$} &  {The system receives the massive disconnecting requests repeatly}\\
      & {$Timer$} &  {System timer that can be used to mark states such as reception frequency and occupancy time}\\
      & {$NonNaviProcess$} &  {The state of non-navigational processes running on the CPU in the system}\\
      & {$AbnormalBehavior$} &  {The state of abnormal behaviors in the system}\\
      & {$NodeID$} &  {Identification of a node in multi-UAV systems}\\
      & {$Authentication$} &  {User permission state of a node in multi-UAV systems}\\
      & {$VideoGet$} &  {The new video data which system receives} \\
      & {$MetaData$} &  {Matching state of attribute information such as metadata in video data}\\
      & {$CrossData$} &  {Matching state of multi-data cross validation in video data}\\
      & {$RiskyCmd$} &  {The system has received an risky control coommand}\\
      & {$RiskyBehavior$} &  {UAV exhibits risky behavior}\\
      & {$GPSRecv$} &  {GPS sensor receives correct data from the satellites}\\
      & {$Direction$} &  {The direction of electronic compass should comply with the GPS data} \\
      & {$Doppler$} &  {The doppler shift of inertial navigation sensor should comply with the GPS data} \\
      \midrule
      \multirow{14}*{Output} & {$UnauthorizedComponent$} &  {Unauthorized component has been detected}\\
      & {$UnauthorizedID$} &  {Unauthorized ID has been detected}\\
      & {$SensorFault$} &  {Sensor fault has been detected}\\
      & {$BUFOverFlow$} &  {Buffer overflow has been detected}\\
      & {$MalInject$} &  {Malicous injectection attack has been detected}\\
      & {$AuthBypass$} &  {Authorization bypass has been detected}\\
      & {$SensorJam$} &  {Sensor jamming has been detected}\\
      & {$APAttack$} &  {Access point attack has been detected}\\
      & {$SigTraBlk$} &  {Signal traffic blocking has been detected}\\
      & {$CtrlCmdSpoofing$} &  {Control commands spoofing has been detected}\\
      & {$SwarmComAttack$} &  {Swarm communication attack has been detected}\\
      & {$SensorSpoofing$} &  {Sensor spoofing has been detected}\\
      & {$ReplayAttack$} &  {Replay attack has been detected}\\
      & {$DangerousClimbRate$} &  {Dangerous climb rate has been detected}\\
      & {$DosAttack$} &  {Dos attack has been detected}\\
      & {$CtrlCmdSpoof$} &  {Control commands spoofing has been detected}\\
      & {$GpsSpoofing$} &  {Sensor spoofing has been detected}\\
      & {$LowBattery$} &  {The battery voltage is low}\\
      & {$OverHeight$} &  {The flight altitude of the UAV is too high}\\
      \midrule
      \multirow{9}*{Countermeasure} & {$RecognizeDeny$} &  {The system refuses to recognize the physical component}\\
      & {$ModifyDeny$} &  {The system rejects this modification}\\
      & {$CmdDeny$} &  {The system refuses to execute the control command}\\
      & {$ProcessInterrupt$} &  {The system interrupts the running process}\\
      & {$RTL$} &  {The system executes the RTL command and return to launch position}\\
      & {$MessageDeny$} &  {The system denies to receive the message}\\
      & {$Hover$} &  {The system executes the HOVER command and hover in the air}\\
      & {$VideoDeny$} &  {The system ground station denies to receive the video data}\\
      & {$Landing$} &  {The UAV lands on the landing field}\\
      &{$EmergLanding$} &  {The UAV encounters emergency and lands on the landing field}\\
      & {$DesendAndDrop$} &  {UAV descends and airdrops the supplies}\\
      \bottomrule
    \end{tabular}
  \end{center}
  \end{table*}

  \subsection{Security Properties of UAS}\label{threats}
   According to the attack vector in \figurename\ {\ref{UAS Security Vulnerabilities}}, we classify the UAS security threats to 3 categories, 13 threats:
  \begin{figure}[htbp]
  \centering
  \includegraphics[width= 9 cm]{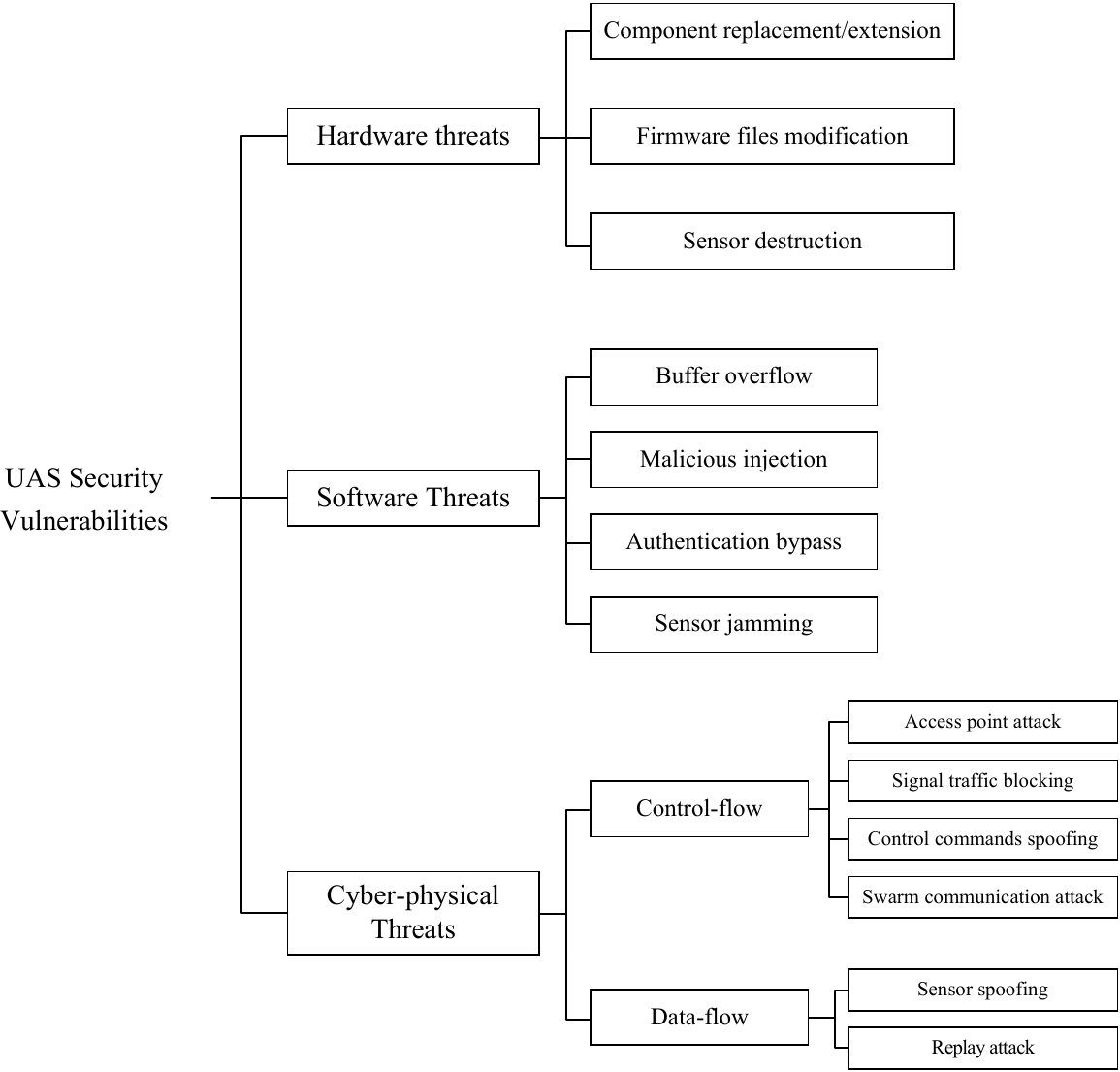}
  \caption{UAS Security Vulnerabilities}\label{UAS Security Vulnerabilities}
  \end{figure}
  
  \subsubsection{Hardware Threats}
Attackers can access to the physical components or local system of UAS, through direct/indirect contact, so as to implement the attacks of component replacement/extension, system code modification/injection or external sensor damage, etc. The damage due to hardware threats depends on the type of ports exposed\cite{7917080}.

1. \textit{Component replacement/extension}. Attackers can connect to UAS and replace some specific components. Malicious components can damage, interfere or even fabricate the internal data and its transmission in the system, so as to affect the performance of the UAS. The device can also be extended by malicious components. For example, during flight network sniffers can be installed on UAV to eavesdrop on and collect the packets in the system which can result in the information leakage.
\begin{table}[!htbp]
  \centering
  
  \caption{Specification Patterns $\mathcal{P}1$: Component replacement/extension}
  \label{propositions1}
  \begin{tabular}{lp{5.5cm}}
  \toprule
  \textbf{Name}& {Fingerprint authorization}\\
  \midrule
  \textbf{Input}& {$ComponentFingerprint$ $TakeOff$ $LandOn$}\\
  \midrule
  \textbf{Property}& \tabincell{l}{The component without correct fingerprint \textbf{shall} \\\textbf{not} be recognized by the system during flight}\\
  \midrule
  \textbf{Specification}& \tabincell{l}{$\Box(TakeOff \Rightarrow (\lnot ComponentFingerprint $ \\$ \ \mathcal{U} \ LandOn))$}\\
  \midrule
  \textbf{Output}& {$UnauthorizedComponent$}\\
  \midrule
  \textbf{Countermeasure}& {$AuthorizeDeny$}\\
  \bottomrule
  \end{tabular}
\end{table}

2. \textit{Firmware files modification}. There are configuration files inside intelligent unmanned systems such as UAS to restrain system in case of illegal behaviors. These files are usually stored statically in the System on Chip (SoC) without any encryption. Attackers can easily access and modify these firmware files to crack the functions for security protection and implement some substantive attacks further.
For example, most commercial UASs would have built-in configuration files about the restrictions of No-fly zone and  height limit. But hackers can modify the configuration files and system parameters, or fabricate the GPS coordinates with cracking module to eliminate the restrictions. Drones without restrictions can fly to anywhere and it is very dangerous.
\begin{table}[!htbp]
  \centering
  
  \caption{Specification Patterns $\mathcal{P}2$: Firmware fles modifcation}
  \label{propositions2}
  \begin{tabular}{lp{5.5cm}}
  \toprule
  \textbf{Name}& {Identity authorization}\\
  \midrule
  \textbf{Input}& {$UserID$ $TakeOff$ $LandOn$}\\
  \midrule
  \textbf{Property}& \tabincell{l}{The firmware files and code \textbf{shall not} be \\modified by unauthorized user during flight.}\\
  \midrule
  \textbf{Specification}& { $\Box(TakeOff \Rightarrow  (\lnot UserID \ \mathcal{U} \ LandOn))$}\\
  \midrule
  \textbf{Output}& { $UnauthorizedID$}\\
  \midrule
  \textbf{Countermeasure}& {$AuthorizeDeny$}\\
  \bottomrule
  \end{tabular}
\end{table}

3. \textit{Sensor destruction}. Light Detection And Ranging (LiDAR) and camera are optical-based external sensors of UAS. They are usually used for the traffic target recognition and distance measurement based on the optical imaging principle. System can accomplish the 3D modeling and display for its surrounding environment based on the data from these sensors, and then make decisions and planning. Jonathan Petit et al.\cite{petit2015remote} present a physical attack on the camera MobilEye C2-270 deployed on an automated vehicle(AV). They emit light of different types into the camera to fully or partially blind it, so as to trigger the system failure of recognizing objects behind the light. He et al.\cite{hedaojing2019} introduce a physical attack on the Inertial Measurement Unit (IMU) of UAS by ultrasonic waves, which is a typical side-channel attack.

\begin{table}[!htbp]
  \centering
  
  \caption{Specification Patterns $\mathcal{P}3$ :Sensor destruction}
  \label{propositions3}
  \begin{tabular}{ll}
  \toprule
  \textbf{Name}& {Sensor faults detection}\\
  \midrule
  \textbf{Input}& {$FusionEstimation$ $TakeOff$ $LandOn$}\\
  \midrule
  \textbf{Property}& \tabincell{l}{The state estimation from the sensor fusion \\algorithm \textbf{shall not} be wrong during flight\cite{liu2013multi}.}\\
  \midrule
  \textbf{Specification}& {$\Box(TakeOff \Rightarrow  (\lnot FusionEstimation \ \mathcal{U} \ LandOn))$}\\
  \midrule
  \textbf{Output}& {$SensorFault$}\\
  \midrule
  \textbf{Countermeasure}& {$RTL$}\\
  \bottomrule
  \end{tabular}
\end{table}

\subsubsection{Software Threats}
Flight control softwares such as Ardupilot and APM, usually have vulnerabilities about not only general software applications but also specific aerial systems. Attackers can exploit these vulnerabilities to implement the attacks of buffer overflow, malware injection, authorization bypass or sensor-data jamming.

4. \textit{Buffer overflow}. Buffer overflow is a very common and dangerous vulnerability for programs. Because there is the existing flaw of no limitation or inspection for the inputs in the target program, attackers can construct some malformed inputs to trigger an anomaly in program. When being written to the buffer, the data will past the boundary of buffer and overwrite adjacent memory locations, this may cause the crash or erratic behaviors of program. Michael Hooper et al.\cite{7795496} implemented a buffer overflow attack on the Parrot Bebop 2. This system utilizes the ARDiscovery protocol to establish a open WI-FI connection between drone and controller(smartphone), and send messages with the JSON records. The Parrot developers seems to have never consider the inspection for the length of each field in the JSON records. So the authors killed the navigational application of the UAV by increasing the character length of the first field in the command sent to the drone, and  the UAV crashed down to the ground immediately.
\begin{table}[!htbp]
  \centering
  
  \caption{Specification Patterns $\mathcal{P}4$ :Buffer overﬂow}
  \label{propositions4}
  \begin{tabular}{lp{5.5cm}}
  \toprule
  \textbf{Name}& {Parameters checking}\\
  \midrule
  \textbf{Input}& {$CmdGet$ $ParmLength$ $TakeOff$ $LandOn$}\\
  \midrule
  \textbf{Property}& \tabincell{l}{The length of each characters in the command \\\textbf{shall not} be over the limit $\lambda_{max}$ during flight.}\\
  \midrule
  \textbf{Specification}& \tabincell{l}{$\Box(TakeOff \Rightarrow  (CmdGet < \lambda_{max} \ \mathcal{U} \ LandOn))$}\\
  \midrule
  \textbf{Output}& {$BUFOverFlow$}\\
  \midrule
  \textbf{Countermeasure}& {$CmdDeny$}\\
  \bottomrule
  \end{tabular}
\end{table}

5. \textit{Malicious injection}. Such attack can aim at UAV and GCS, respectively: \textbf{Injection to GCS} can be the programs such as virus, trojan and malware, etc., injected into the computer or other control devices. When running in the system, it will the trigger the anomaly, data leakage or lossing control. \textbf{Injection to UAV} can be mainly the control commands. Due to the logic vulnerabilities, some control softwares fail to verify the validation of the commands and get hijacked.
\begin{table}[!htbp]
  \centering
  
  \caption{Specification Patterns $\mathcal{P}5$ :Malicious injection}
  \label{propositions5}
  \begin{tabular}{lp{5.5cm}}
  \toprule
  \textbf{Name}& {System performance monitoring}\\
  \midrule
  \textbf{Input}& {$CPUuse$ $MEMuse$ $I_{batt}$ $TakeOff$ $LandOn$}\\
  \midrule
  \textbf{Property}& \tabincell{l}{The utilization of CPU and memory \textbf{shall not} \\more than $\alpha$\%, and battery current \textbf{shall not} \\be more than n A for 10 time steps during flight.}\\
  \midrule
  \textbf{Specification}& \tabincell{l}{$\Box(TakeOff \Rightarrow ((CPUuse < a\% \land  MEMuse $ \\$< a\%  \land \lnot \Box_{[0, 10]} I_{batt} > n A) \ \mathcal{U} \ LandOn))$}\\
  \midrule
  \textbf{Output}& {$MalInject$}\\
  \midrule
  \textbf{Countermeasure}& {$ProcessInterrupt$}\\
  \bottomrule
  \end{tabular}
\end{table}

6. \textit{Authorization bypass}. As one of the Internet of Things(IoT), UAS support numbers of different interfaces and services for users. The authorization of the identity becomes particularly important for the system. Authorization can prevent the invalid access, because only the authorized users are allowed to access to the system with the appropriate permissions. But lots of UASs do not perform well in the protection of access permission. None or weak protection will help attackers to bypass the authorization and control the system directly. Rahul Sasi\cite{sasi2015maldrone} showed his drone backdoor for the Parrot AR Drone 2 in 2015. It can disconnect the GCS from UAV and establish a reverse TCP connection between the UAV and fake GCS(attacker), to actually control  and hijack the UAV.
\begin{table}[!htbp]
  \centering
  
  \caption{Specification Patterns $\mathcal{P}6$ : Authorization bypass}
  \label{propositions6}
  \begin{tabular}{ll}
  \toprule
  \textbf{Name}& {Command verification}\\
  \midrule
  \textbf{Input}& {$Mode$ $AbnormalCmd$ $TakeOff$ $LandOn$}\\
  \midrule
  \textbf{Property}& \tabincell{l}{UAV \textbf{shall not} execute any abnormal commands \\in a specific mode during flight.}\\
  \midrule
  \textbf{Specification}& {$\Box(TakeOff \Rightarrow  (\lnot AbnormalCmd \ \mathcal{U} \ LandOn))$}\\
  \midrule
  \textbf{Output}& {$AuthBypass$}\\
  \midrule
  \textbf{Countermeasure}& {$CmdDeny$}\\
  \bottomrule
  \end{tabular}
\end{table}

7. \textit{Sensor jamming}. Data from the sensors is always correct by default, but it is not true in fact. Although manufacturers can verify the robustness and resilience of the environment modeling with the mechanisms of sensors fusion and redundancy, there is still a attack surface exposed to the hackers. Alan Kim et al.\cite{kim2012cyber} introduce a potential attack: attackers can interfere the analog signals or change the sampling rate of the digital signals that can generate "dirty data" to influence the system performance.
\begin{table}[!htbp]
  \centering
  
  \caption{Specification Patterns $\mathcal{P}7$ :Sensor jamming}
  \label{propositions7}
  \begin{tabular}{ll}
  \toprule
  \textbf{Name}& {Sensor jamming detection}\\
  \midrule
  \textbf{Input}& {$FusionEstimation$ $TakeOff$ $LandOn$}\\
  \midrule
  \textbf{Property}& \tabincell{l}{The sensor error \textbf{shall not} last more than \\ $n$ time steps during flight.}\\
  \midrule
  \textbf{Specification}& \tabincell{l}{$\Box(TakeOff \Rightarrow  ((SensorError \ \mathcal{U}_{[0,n]} $\\$SensorData )) \ \mathcal{U} \ LandOn))$}\\
  \midrule
  \textbf{Output}& {$SensorJam$}\\
  \midrule
  \textbf{Countermeasure}& {$RTL$}\\
  \bottomrule
  \end{tabular}
\end{table}
\subsubsection{Cyber-Physical Threats}
UAS communication can be radio communication, wireless network communication and satellite communication according to the bandwith. The radio communication and wireless network communication are within the visual range(WVR), and satellite communication is beyond the visual range(BVR). Information transmission of UAS can be \textbf{control-flow} and \textbf{data-flow}. For control-flow, attackers can implement the attack of access-point(AP) intrusion, signal traffic-block, control-signal spoofing. For data-flow, attackers can implement the sensor-data spoofing attack and replay attack.

\begin{itemize}
\item \textbf{Control-flow}
\end{itemize}

8. \textit{Access point attack}. Typical access point attacks include man-in-the-middle(MITM) attack and Deauthentication(De-Auth) attack. Most communication protocols for UAS(MAVLink, for instance) are designed without sufficient security protection.
\textbf{MITM attack} connects to the network and intercepts the packets between GCS and UAV via the ARP poisoning or DNS spoofing, and then resend the tampered packets to the APs. Attacked system has no awareness of the attacker at all in the meantime.
\textbf{De-Auth attack} is one of DoS attacks and aims at sending the massive disconnecting requests repeatly. If it works, attackers can connect to the AP and control the system. Samy Kamkar\cite{dey2018security} successfully implemented a De-Auth attack on the AR.Drone 2 with his drone-hijacking software called SkyJack in 2013.
\begin{table}[!htbp]
  \centering
  
  \caption{Specification Patterns $\mathcal{P}8$ :Access point attack}
  \label{propositions8}
  \begin{tabular}{lp{5.5cm}}
  \toprule
  \textbf{Name}& {Command checking}\\
  \midrule
  \textbf{Input}& {$TakeOff$ $DeauthCmd$ $TakeOff$ $LandOn$}\\
  \midrule
  \textbf{Property}& \tabincell{l}{UAV \textbf{shall not} receive the De-auth command \\ for at most 5 time steps during flight.}\\
  \midrule
  \textbf{Specification}& \tabincell{l}{$\Box(TakeOff \Rightarrow  (\lnot \Box_{[0, 5]} DeauthCmd \ \mathcal{U} $\\$\ LandOn))$}\\
  \midrule
  \textbf{Output}& \tabincell{l}{$APAttack$}\\
  \midrule
  \textbf{Countermeasure}& {$CmdDeny$}\\
  \bottomrule
  \end{tabular}
\end{table}

 9. \textit{Signal Traffc Blocking}. Signal analysis and processing is restricted to the limited resources in the embedded systems such as UAS. Massive and incessant data sent to the UAS will easily overwhelm and exhaust the resources.
DoS or DDoS attack for UAS is fatal and hard to defend against. Because the packets sent to the UAS are valid and can consume too much resources, system will get occupied and fail to respond to the normal requests. Gabriel Vasconcelos et al.\cite{vasconcelos2016impact} perform an experiment to evaluate the impact of DoS attacks on the AR.Drone 2.0 with three attack tools: LOIC, Netwox, and Hping3.
Directional radio frequency interference is a simple way to jamming the target UAS by sending some signals with specific direction, power, and frequency.

\begin{table}[htbp]
  \centering
  \caption{Specification pattern $\mathcal{P}9$ :Signal traffc blocking}
  \label{dosatk}
  \begin{tabular}{lp{5.5cm}}
  \toprule
  \textbf{Name}& {Process monitoring}\\
  \midrule
  \textbf{Input}& {$Guided$ $AbnormalCmd$ $CPUuse$}\\
  \midrule
  \textbf{Property}& \tabincell{l}{UAV \textbf{shall not} receive abnormal commands \\ for at most 10 time steps and the CPU usage of \\  this kind of command \textbf{shall not} always be over \\ $\alpha \%$ for at most 5 time steps in GUIDED mode.}\\
  \midrule
  \textbf{Specification}& \tabincell{l}{$\Box(Guided \Rightarrow (\lnot \Box_{[0, 5]} AbnormalCmd) \land $\\$ (AbnormalCmd \Rightarrow \Diamond_{[0,5]} CPUuse < \alpha \%  )))$}\\
  \midrule
  \textbf{Output}& {$DosAttack$}\\
  \midrule
  \textbf{Countermeasure}& {$CmdDeny$}\\
  \bottomrule
  \end{tabular}
\end{table}

    

10. \textit{Control Commands Spoofing}. As shown in Fig. \ref{spfatk}, this attack often occurs due to existing logic flaws or weak authorization. For example, the DJI Phantom \uppercase\expandafter{\romannumeral3} has been totally hijacked by 
hackers in the GeekPwn 2015. Hackers cracked the signals of the chip BK5811 boarded on the RC controller of Phantom 3, and then exploited the chip's design vulnerabilities to generate spurious commands.

\begin{table}[htbp]
  \centering
  \caption{Specification pattern $\mathcal{P}10$ :Control commands spoofing}
  \label{spfatk}
  \begin{tabular}{lp{5.5cm}}
  \toprule
  \textbf{Name}& {Risky behaviors detection}\\
  \midrule
  \textbf{Input}& {$Guided$ $RiskyCmd$ $RiskyBehavior$}\\
  \midrule
  \textbf{Property}& \tabincell{l}{Risky behavior \textbf{shall not} occur in 5 time \\steps after UAV receive a risky command \\in Guided mode.}\\
  \midrule
  \textbf{Specification}& \tabincell{l}{$\Box(Guided \land (RiskyCmd \Rightarrow \Diamond_{[0,5]} $\\$ \lnot RiskyBehavior) )$}\\
  \midrule
  \textbf{Output}& {$CtrlCmdSpoofing$}\\
  \midrule
  \textbf{Countermeasure}& {$RTL$}\\
  \bottomrule
  \end{tabular}
\end{table}

11. \textit{Swarm communication attack}. The most security issue for the multi-UAV systems such as formation and swarm is the security of Flying Ad-hoc Networks(FANETs). Communication link in the FANETs is essential and complicated, and the heterogeneous networks between different subsystems may also bring more security vulnerabilities. Ilker Bekmezci et al.\cite{bekmezci2016security} made a comprehensive study on security of FANETs in 2016, and Nicolas Lechevin et al.\cite{lechevin2015health} introduce the security threats from the physical layer, data-link layer and network layer of multi-UAV systems.
\begin{table}[!htbp]
  \centering
  
  \caption{Specification Patterns $\mathcal{P}11$ :Swarm communication attack}
  \label{propositions11}
  \begin{tabular}{lp{5.5cm}}
  \toprule
  \textbf{Name}& {Mutual authentication}\\
  \midrule
  \textbf{Input}& {$TakeOff$ $MalMessage$ $LandOn$}\\
  \midrule
  \textbf{Property}& \tabincell{l}{The message from the node marked as \\malicious node \textbf{shall not} be accepted\cite{FANET2018} during \\flight.}\\
  \midrule
  \textbf{Specification}& \tabincell{l}{$\Box(TakeOff \Rightarrow  (\lnot MalMessage \ \mathcal{U} \ LandOn))$}\\
  \midrule
  \textbf{Output}& {$SwarmComAttack$}\\
  \midrule
  \textbf{Countermeasure}& {$MessageDeny$}\\
  \midrule
  \end{tabular}
\end{table}

\begin{itemize}
\item \textbf{Data-flow}
\end{itemize}

   12.  \textit{Sensor Spoofing} Data from sensors is the most important input for the UAS. Besides the control-flow, attackers can also interfere with or spoof sensors data to affect the system performance.
For example, GPS spoofing is the most common attack and can be autonomous spoofing and forwarding spoofing. Many cases are studied in \cite{8088163}.
 Countermeasures could be the mechanism of verifying the difference between multiple sensors about (1) motion-speed (e.g., the coordinates of UAV cannot change from Beijing to NewYork in a few seconds), 
 and (2)time of GPS-time and NTP-time \cite{dey2018security} (3)directions of GPS and electronic compass, etc. 

\begin{table}[htbp]
  \centering
  \caption{Specification pattern $\mathcal{P}12$ :Sensor spoofing}
  \label{senspfatk}
  \begin{tabular}{lp{5.5cm}}
  \toprule
  \textbf{Name}& {Sensor faults detection}\\
  \midrule
  \textbf{Input}& {$TakeOff$ $GPSRecv$ $SubFrame$ $Direction$ $Doppler$ $LandOn$}\\
  \midrule
  \textbf{Property}& \tabincell{l}{During flight, a GPS signal received \textbf{shall} keep \\ consistent with the results of arrival direction \\and doppler shift of inertial navigation sensor}\\
  \midrule
  \textbf{Specification}& \tabincell{l}{$\Box(TakeOff \Rightarrow  (GPSRecv \land Direction $\\ $ \land Doppler)\ \mathcal{U} \  LandOn))$}\\
  \midrule
  \textbf{Output}& {$GPSSpoofing$}\\
  \midrule
  \textbf{Countermeasure}& {$Hover$}\\
  \bottomrule
  \end{tabular}
\end{table}

13. \textit{Replay attack}. Video feedback from UAV to GCS can be transmitted in several ways. Attackers usually intercept, modify and forward the video data to confuse the GCS. A presentation by Aaron Luo at DEFCON 2016 portrayed a video replay attack to DJI Phantom \uppercase\expandafter{\romannumeral3}, that the video displayed on the screen of mobile controller was replaced by a picture.
\begin{table}[!htbp]
  \centering
  
  \caption{Specification Patterns $\mathcal{P}13$ :Replay attack}
  \label{propositions13}
  \begin{tabular}{lp{5.5cm}}
  \toprule
  \textbf{Name}& {Information validation}\\
  \midrule
  \textbf{Input}& {$VideoGet$ $MetaData$ $CrossData$}\\
  \midrule
  \textbf{Property}& {GCS \textbf{shall not} accept the wrong video data.}\\
  \midrule
  \textbf{Specification}& \tabincell{l}{$\Box(VideoGet \land (\lnot MetaData \lor \lnot CrossData))$}\\
  \midrule
  \textbf{Output}& {$ReplayAttack$}\\
  \midrule
  \textbf{Countermeasure}& {$VideoDeny$}\\
  \midrule
  \end{tabular}
\end{table}

\subsubsection{Safety properties of UAS}
Unlike security threats that mainly caused by attackers, safety threats are mainly from the environment and UAV itself, such as battery consumption, maximum flight speed, system memory, CPU occupancy and obstacle avoiding etc. 
For example, to prevent UAV from being unable to return due to low battery, we can set the minimum battery level of the UAV to 50 \% during flight.

\begin{table}[htbp]
  \centering
  \caption{Specification pattern $\mathcal{P}14$ :Safety constraint}
  \label{safepat}
  \begin{tabular}{ll}
  \toprule
  \textbf{Name}& {Low Battery}\\
  \midrule
  \textbf{Input}& {$Takeoff$ $Battery$}\\
  \midrule
  \textbf{Property}& \tabincell{l}{During flight, the minimum battery level \textbf{shall} \\ \textbf{not} be less than 50 \% .}\\
  \midrule
  \textbf{Specification}& {$\Box(Takeoff \Rightarrow  (Battery > 50 \% ))$}\\
  \midrule
  \textbf{Output}& {$LowBattery$}\\
  \midrule
  \textbf{Countermeasure}& {$RTL$}\\
  \bottomrule
  \end{tabular}
\end{table}

\subsubsection{Task properties of UAS}
We can also use our UAV monitor property pattern to design some simple tasks in the form of $Condition \rightarrow Action$ (i.e., some action should be taken under certain condition).
For example, UAV airdrops to the injured at high altitude may cause supplies to deviate from the target point, thus UAV height shall not be over 30 m when finding the injured. 
\begin{table}[htbp]
  \centering
  \caption{Specification pattern $\mathcal{P}15$ :Task constraint}
  \label{taskpat}
  \begin{tabular}{ll}
  \toprule
  \textbf{Name}& {Minimum altitude for UAV airdrop}\\
  \midrule
  \textbf{Input}& {$Injured$ $Flying$ $Height$}\\
  \midrule
  \textbf{Property}& \tabincell{l}{The UAV height \textbf{shall not} be over 30 m\\ when finding the injured in the air.}\\
  \midrule
  \textbf{Specification}& {$\Box(Flying \land Injured \Rightarrow  (Height < 30 m))$}\\
  \midrule
  \textbf{Output}& {$OverHeight$}\\
  \midrule
  \textbf{Countermeasure}& {$DesendAndDrop$}\\
  \bottomrule
  \end{tabular}
\end{table}



\section{Secure Controller for UAS}\label{Construction}
  UAS consists of UAV and GCS, so we consider UAV as robot and GCS as environment in this paper. The communication between UAV and GCS can be abstracted into the process of robot sensing data from its environment.
  
  For \textit{Positions} modeling, in this paper we specify the workspace in \figurename{\ref{map}}. It is shown that there are 6 zones partitioned in the workspace: \textit{StartZone}, $p_1$, $p_2$, $p_3$, $p_4$ and $p_5$.
  For \textit{Actions} modeling, in order to increase the reusability of modules, we propose to separate the functionality and security of UAV and divide UAV actions into two components: \textit{actuator} and \textit{monitor}, which are loosely coupled.
  \textit{Actuator} focuses on satisfying the functional requirements of UAS, by executing the dynamic actions. We assume that UAV has 5 action modes mapping to the corresponding real flight operations of UAV.
  \textit{Monitor} mainly verifies the real-time execution of UAV satisfying the given security properties or not, by analyzing the specific data. So we assume that UAV has a action mode 
  MONITOR. The function of these action modes are shown in \tablename\ {\ref{mode}}.
  For \textit{Sensors} modeling, the sensors of \textit{monitor} and \textit{actuator} would be discussed respectively. Here We take the control commands spoofing attack as an example:
  
  \begin{figure}[htbp]
    \centering
    \includegraphics[width=2.8in]{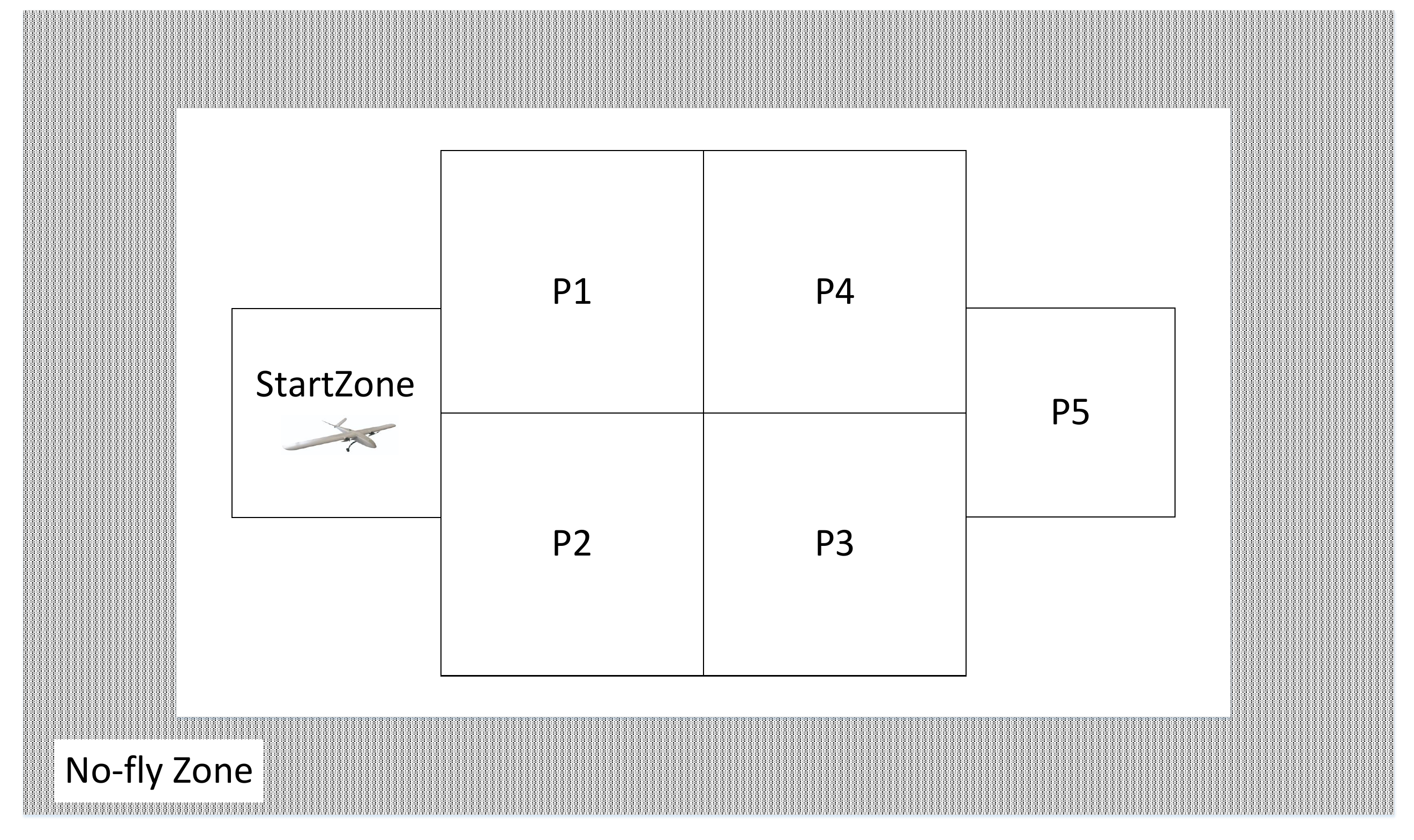}
    \caption{The workspace of UAS model}\label{map}
    \end{figure}

  \textbf{Example 1}:UAV takes off from \textit{StartZone} and patrols from $R1$ to $R5$ in sequence as shown in \figurename{\ref{map}}. We extend the specification pattern 10 as 
  shown in R. When abnormal instructions such as modifying parameters or clearing waypoints are sent in specific flight modes 
  and cause the abnormal behavior of the system, this scenario can be regarded as an control commands spoofing attack. In order to detect this attack, we monitor the flight coordinate range and other data of 
  the UAS in AUTO mode as inputs. If the UAV receives the waypoints-clearing command and replan the waypoints, resulting in the flight trajectory deviating from the 
  specified area into the no-fly zone, it can be considered that the UAV are under the control commands spoofing attack.

  Countermeasures could be flight mode changing to RTL mode. We can describe the security property \textbf{R} and LTL specification like this:
  \begin{itemize}
    \item [\textbf{R}:]During flight, if UAV is in AUTO mode, receives waypoints-clearing commands, replans new waypoints and  flys into the no-fly zone, it encounters control commands spoofing attack.
  \end{itemize}
  $$\Box((\textit{WPclearcmd}\land \textit{armed}\land \textit{mode\_auto}\land \textit{NewWP}\land \textit{NoflyZone})$$
  $$ \Rightarrow \bigcirc \textit{WPclearAttack})$$

  \renewcommand\arraystretch{1}
  \begin{table}[htbp]
  \centering
  \caption{Action Modes of UAV}\label{mode}
  \scalebox{1}[1]{
  \begin{tabular}{cl}
  \toprule
  \textbf{Modes}& \textbf{Function Description}\\
  \midrule
  TAKEOFF& UAV must take off from the zone \textit{StartZone}\\
  LAND& UAV must land at the zone \textit{StartZone}\\
  \multirow{2}*{AUTO}& UAV must cruise around the workspace in order \\
  & from $p_1$ to $p_5$\\
  RTL& UAV must return to the launch site(\textit{StartZone}) \\
  \multirow{2}*{MONITOR}& When detecting a property violation, \textit{monitor} must \\
  & force \textit{actuator} to  defend against the attack\\
  \bottomrule
  \end{tabular}}
  \end{table}

  \begin{figure*}[htbp]
    \centering
    \centerline{\includegraphics[width=0.7\textwidth]{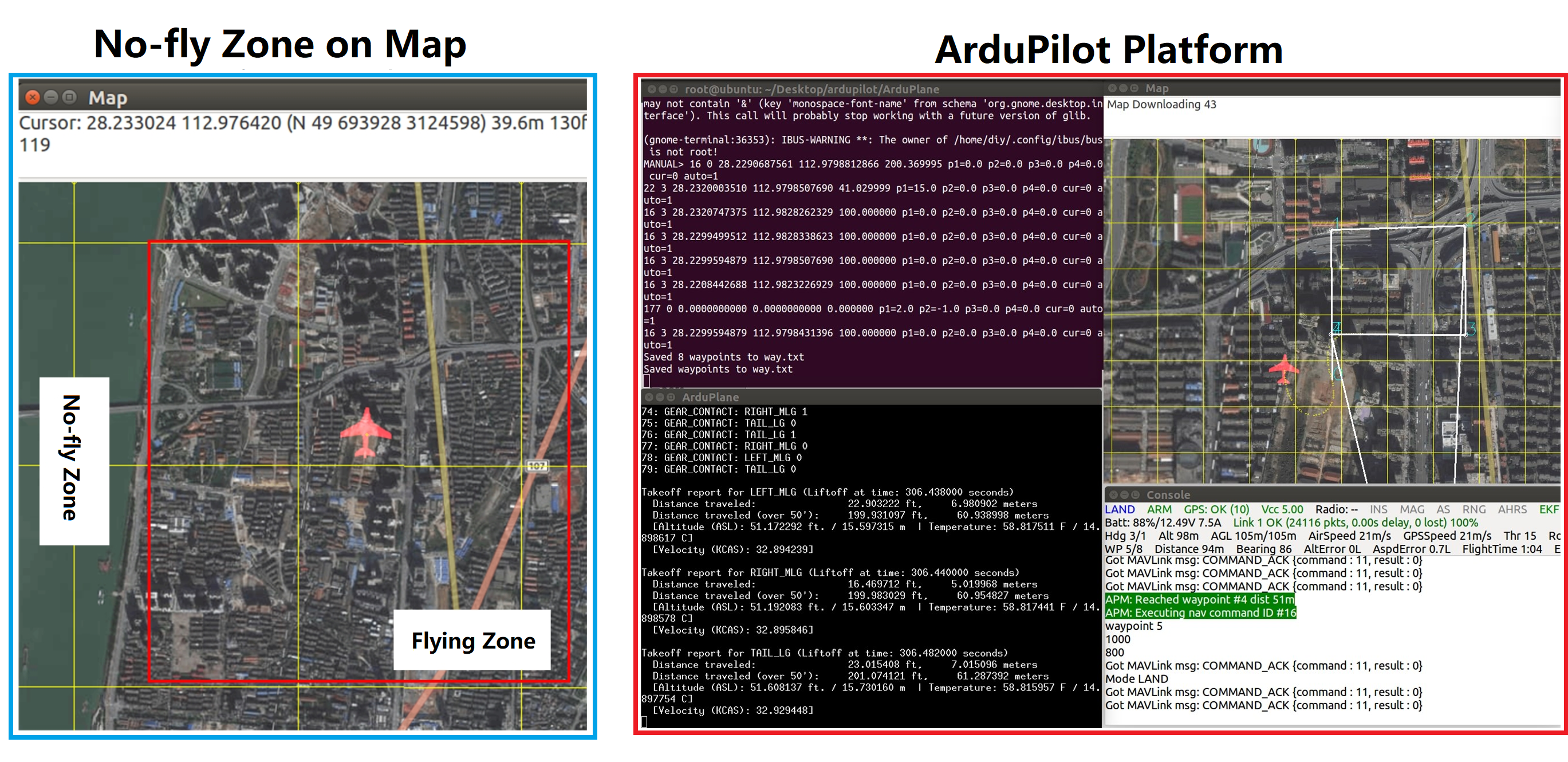}}
    \caption{No-fly zone and simulation of UAS}\label{exp1}
    \end{figure*}

  \begin{figure*}[htbp]
    \centering
    \centerline{\includegraphics[width=0.7\textwidth]{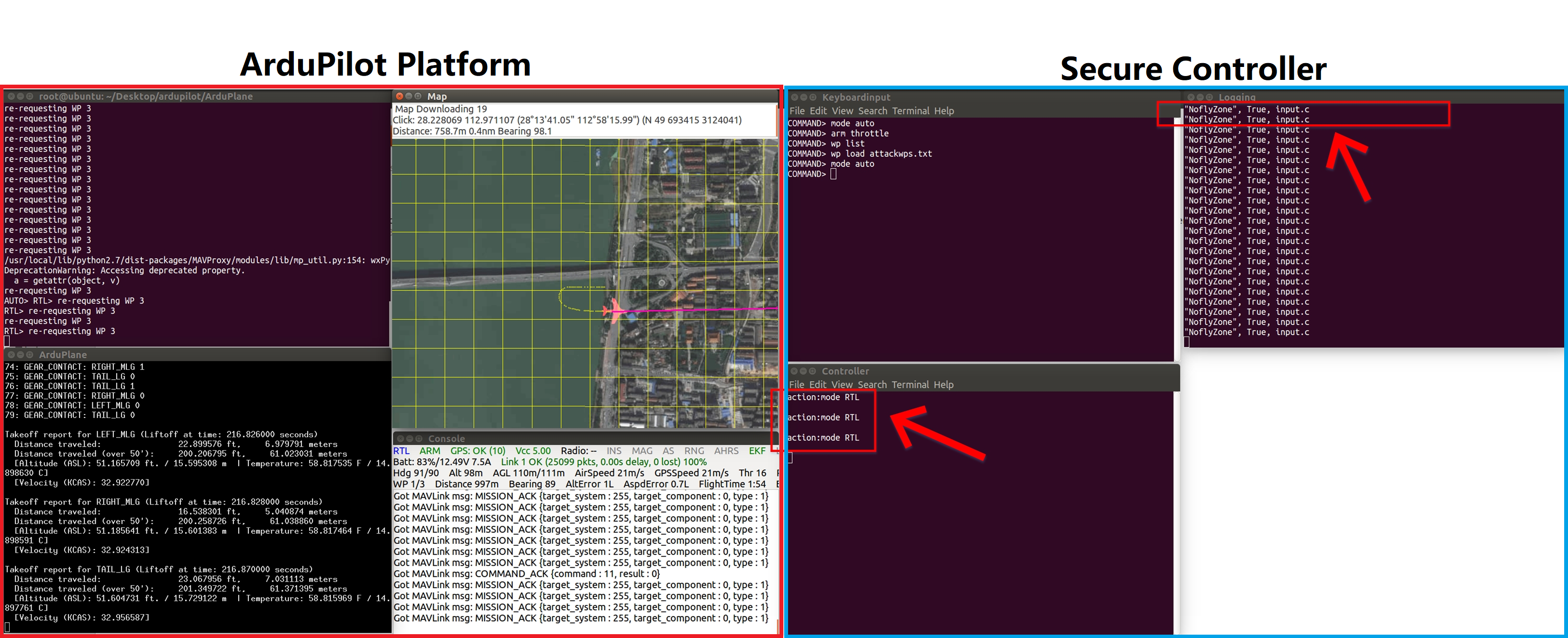}}
    \caption{The simulation of secured UAS model}\label{exp2}
    \end{figure*}

  For \textbf{Example 1},we should first define propositions sets are shown in \tablename\ {\ref{propositions}}, then we can we describe the motion planning and security specification \textbf{R} of \textit{actuator} 
  and \textit{monitor} as follows, detailed specifications are given in Fig. \ref{spec4UAV}. Each component of UAV can be seen as part of the other one's environment, and obtain the dynamic information from its sensors to achieve the purpose of communication. 
  
  \renewcommand\arraystretch{1}
  \begin{table}[htbp]
  \centering
  \caption{Propositions Set for UAS Model}
  \label{propositions}
  \scalebox{1}[1]{
  \begin{tabular}{c|l}
  \hline
  \textbf{Set}& \textbf{Propositions}\\
  \hline
  $\mathcal{X}_a$& $\textit{TAKEOFF}, \textit{AUTO}, \textit{Mnt\_WPclearAttack}$\\
  \hline
  $\mathcal{X}_m$& $\textit{Act\_armed}, \textit{Act\_mode\_auto},\textit{WPclearcmd}, \textit{NewWP},\textit{NoflyZone}$\\
  \hline
  $\mathcal{Y}_a$& $\textit{StartZone}, p_1, \ldots, p_5,\textit{armed}, \textit{mode\_auto}, \textit{mode\_RTL}$\\
  \hline
  $\mathcal{Y}_m$& $\textit{WPclearAttack}$\\
  \hline
  \end{tabular}}
  \end{table}

  \begin{enumerate}
    \item \textit{Actuator} cannot be in \textit{auto} and \textit{RTL} mode simultaneously.
    \item \textit{Actuator} must arm throttle when it receives the \textit{TAKEOFF} command and is not armed.
    \item \textit{Actuator} must change to \textit{auto} mode when it receives the \textit{AUTO} command and is not armed in fly zone.
    \item When sensing the alarm \textit{WPclearAttack} from \textit{monitor}, \textit{actuator} must turn to mode \textit{RTL}.
    \item \textit{Monitor} always monitors the system. If a flying UAV in \textit{auto} mode receives waypoints-clearing command, changes waypoints and enters no-fly zone, \textit{monitor} must 
    send \textit{WPclearAttack}  alarm and force \textit{actuator} to defend against the attack.
  \end{enumerate}

  Given the motion planning and security specification mentioned above, automatons for the controller of \textit{actuator} and \textit{monitor} will be generated respectively. These components implement the game 
  process with environment coordinately to find paths satisfying the goal $\phi$ no matter how the environment changes within the $\varphi^e_t$.
  We finally implement the translation from high-level requirements to the practical automatons with specific $\varphi_i$, $\varphi_t$ and $\varphi_g$. Details will be discussed in Section \ref{implement}.

  \section{Implement And Experiment} \label{implement}
  
  We implement the tasks translation and automatons generation for the given motion planning and security requirement in \textbf{Example 1} with GR(1) synthesis algorithm. Here we denote \textit{actuator} by $\varphi^a$ and \textit{monitor} by $\varphi^m$.

  As shown in Tab. \ref{spec4UAV}, first of all, we should specify the initial states of UAS ($\varphi^{s,a}_i$ and $\varphi^{s,m}_i$)   and its environment ($\varphi^{e}_i$).Then the planning are translated into $\varphi_t$
  Finally, the goal of UAV is flying according to the action mode, and always satisfy the security properties. This requirement is translated into $\varphi^{s}_g$.

  \begin{table}[!htbp]
    \begin{center}
      \caption{Specifications for UAV.}
      \label{spec4UAV}
      \scalebox{1}{
      \begin{tabular}{cl}
        \toprule
        \textbf{Categories} & \multicolumn{1}{c}{\textbf{Specifications}}\\
      \toprule
      $\varphi^{e}_i$ & \tabincell{l}{$\lnot\textit{TAKEOFF}, \lnot\textit{armed}, \lnot\textit{AUTO}, \lnot\textit{WPclearcmd},$\\$ \lnot\textit{NewWP},  \lnot\textit{NoflyZone},\lnot\textit{Mnt\_WPclearAttack}$}\\
      \cmidrule{2-2}
        $\varphi^{s,a}_i$ & $\textit{StartZone}, \lnot\textit{armed}, \lnot\textit{mode\_auto}, \lnot\textit{mode\_RTL}$\\
        \cmidrule{2-2}
        $\varphi^{s,m}_i$ & $\lnot WPclearAttack$\\
        \midrule
        \multirow{4}*{$\varphi^{s,a}_t$} & $\Box \lnot (\textit{mode\_auto}\land \textit{mode\_RTL}),$ \\ & $\Box((\textit{TAKEOFF}\land \lnot\textit{armed})\Rightarrow \bigcirc \textit{armed}),$ \\ & $\Box ((\textit{AUTO}\land\textit{armed}\land\lnot\textit{NoflyZone})\Rightarrow \bigcirc \textit{mode\_auto}),$ \\ & $\Box(\textit{Mnt\_WPclearAttack}\Rightarrow \bigcirc \textit{mode\_RTL})$\\
        \cmidrule{2-2}
        $\varphi^{s,m}_t$ & \tabincell{l}{$\Box((\textit{WPclearcmd}\land \textit{Act\_armed}\land \textit{Act\_mode\_auto}\land $\\$  \textit{NewWP}\land \textit{NoflyZone})\Rightarrow \bigcirc \textit{WPclearAttack})$}\\
        \midrule
        \multirow{1}*{$\varphi_g$} & $\Box\Diamond(\textit{P1}),\,\,\Box\Diamond(\textit{P2}),\,\,\Box\Diamond(\textit{P3}),\,\,\Box\Diamond(\textit{P4}),\,\,\Box\Diamond(\textit{P5})$\\
        \bottomrule
      \end{tabular}}
    \end{center}
    \end{table}

  We test the effectiveness of generated automatons  in \textit{Example 1}  on Ardupilot platform and evaluate the result as follows:

  \figurename{\ref{exp1}} shows the no-fly zone on the map and  console screen of simulation recording the status of UAV during the experiment, the UAV is taking off from  \textit{StartZone} and patrolling from $P_1$ to$ P_5$.

  In \figurename{\ref{exp2}}, we simulate the waypoints clearing attack and lead the UAV to enter the no-fly zone. Our secure controller detects this attack and forces the UAV to return to the launch position.

  In this experiment, we automatically generate a robot controller in the form of automaton with LTLMoP platform. Of course it works well in other modes and attack scenarios, too. This mechanism undoubtedly ensures 
  UAV running in a given environment and protect it from malicious attacks.
  
  \section{Related Work}
  In recent years, commercial drones and smart cars suffered from many malicious attacks and resulted in several traffic accidents. So the core issue in this paper is the automatic generation of UAS security behaviors,
   and study on UAS vulnerabilities and corresponding attack chains become essential and fundamental.
  Most UASs consist of several components: control unit, dynamics module, navigation module, sensor module, communication module and so on. The work in \cite{6459914} introduced Confidentiality, Integrity and 
  Availability from the perspective of information security. Confidentiality means that the system forbids illegal accessing or interception of data; Integrity is the property that protects the system from jamming 
  by malicious data; Availability refers to a timely response for legitimate requests. 
  Vrizlynn\cite{7917080} and Leela\cite{8088163} divided attack vector into physical-attack and remote-attack. Regarding this, 
  UAS architecture can be divided into 4 layers based on CPS architecture: sensing layer, execution layer, control layer and data transmission layer. So we combine the security issues with UAS architecture, and 
  classify UAS security vulnerabilities to the categories of hardware, software and Cyber-physical layer.
  
  Runtime verification is a technology that checks if an execution of system satisfying the given properties\cite{j.jlap.2008.08.004}. The process logging and analyzing the finite trajectory of the program and 
  verifies whether it satisfies the properties, with a monitor of properties. The monitor can be online or offline, and constructed by the conversion algorithm which converts a specification describing properties 
  into an equivalent automaton.
  In \cite{Moosbrugger2017}, the authors implement the RV monitor on UAS using LTL formulas to describe the security requirements, with Bayesian Networks and FPGA hardware platform.
  
  Reactive Synthesis for logical specifications is considered a challenging problem\cite{DBLP:journals/jcss/BloemJPPS12} and got effective progress with the development of Temporal Logic and especially 
  LTL, translating the specifications to automaton.
  Due to the double-exponential time complexity, Roderick Bloem\cite{DBLP:journals/jcss/BloemJPPS12} proposed a game-based synthesis algorithm with GR(1) formulas, 
  which address the problem of exponential explosion by segmenting LTL specifications into the sets of \textit{assumptions} and \textit{guarantees}. Based on GR(1) games, the algorithm can extract a winning 
  strategy for the system.  After the synthesis of GR(1) formula is proposed, it is applied in many fields, such as synthesis for specification patterns \cite{Maoz2015GR}, executable PLC code synthesis, 
  controller synthesis for robotics \cite{2020Iterative,2020Controller}, etc. 
  
  Our work in this paper draws on the theory about RV, and constructs a model of UAS with actuator and runtime monitor by the reactive synthesis algorithm.
  
  \section{Conclusion}
  In this paper, we have surveyed security vulnerabilities of UAS associated with the hardware, software and cyber-physical layer, and introduce a specification patterns-based framework to extract the security properties for the threats with LTL formulas. Then we propose an approach based on reactive synthesis to construct a secure UAS controller automatically. GPS attack is one of the most severe threats in UAS applications, unencrypted data and unsecured transmission incurring lots of troubles. By abstracting a model of UAS, we constructing a controller executing specific motions and a runtime monitor detecting the property violations. The model described by LTL specifications can be translated to a strategy automaton by synthesis algorithm. The automatic generation is intuitively a two-player game between UAV and environment, and proved to be an effective method to protect UAS from attacks such as GPS spoofing.
  
  In the future, we are planning to use the state-of-the-art Spectra tools \cite{2019Spectra} to design more complex UAV tasks and analyze more security requirements, and make an implementation on the real UAS platform such as \textit{Ardupilot}.

\IEEEdisplaynontitleabstractindextext

%
\IEEEpeerreviewmaketitle

\ifCLASSOPTIONcaptionsoff
  \newpage
\fi



%

\bibliographystyle{IEEEtran}
\bibliography{ref}

\end{document}